\documentclass[12pt]{article}
\usepackage{authblk}
\usepackage{graphicx}

\begin{document}

\title{\bf On Quantum Mechanics Fundamentals of Diatomic Molecular Spectroscopy}

\author{C. G. Parigger} 
\affil{Physics and Astronomy Department, University of Tennessee, \\ 
University of Tennessee Space Institute, \\ Center for Laser Applications, \\
411 B.H. Goethert Parkway, \\ Tullahoma, TN 37388-9700, USA; \\
cparigge@tennessee.edu; Tel.: +1-(931)-841-5690}
\date{}
\maketitle

{\noindent Keywords: Foundations of Quantum Mechanics, Molecular Spectroscopy, Diatomic Molecules, Symmetry Transformations, Optical Emission Spectroscopy, Astrophysics}


\begin{abstract}
The interpretation of optical spectra requires thorough comprehension of quantum mechanics, especially understanding the concept of angular momentum operators. Suppose now that a transformation from laboratory-fixed to molecule-attached coordinates, by invoking the correspondence principle, induces reversed angular momentum operator identities. However, the foundations of quantum mechanics and the mathematical implementation of specific symmetries assert that reversal of motion or time reversal includes complex conjugation as part of anti-unitary operation. Quantum theory contraindicates sign changes of the fundamental angular momentum algebra. Reversed angular momentum sign changes are of a heuristic nature and are actually not needed in analysis of diatomic spectra. This review addresses sustenance of usual angular momentum theory, including presentation of straightforward proofs leading to falsification of the occurrence of reversed angular momentum identities. This review also summarizes aspects of a consistent implementation of quantum mechanics for spectroscopy with selected diatomic molecules of interest in astrophysics and in engineering applications.
\end{abstract}


\section{Introduction}

Identification of diatomic molecular spectra necessitates a clear description of angular momentum (AM) in order to demarcate the various features that comprise optical fingerprints. Quantum mechanics theory (QMT) asserts that not all three components of AM can be measured simultaneously, usually the total AM and one projection of the total AM describe upper and lower states of molecular transitions.


Classical mechanics (CM) description and associated quantization of the asymmetric top \cite{OKlein} suggests occurrence of commutator relations with different signs when computing momenta with respect to the principal axes of inertia. In other words, a laboratory-fixed system shows standard AM commutators, but with respect to the molecule-attached coordinate system, there is a sign change that carries the name ``reversed'' internal AM~\cite{VanVleck}. The derivation by Klein in 1929 \cite{OKlein} is based on the correspondence principle that in essence emphasizes that QMT reproduces classical physics in the limit of large quantum numbers. From a CM point of view, reversal of motion occurs when transforming from a lab-fixed to a molecule-attached coordinate system, akin to experience of motion reversal when jumping onto a moving merry-go-around. However, reversal of motion in quantum mechanics (QM) is described by an anti-unitary transformation, requiring sign change and complex conjugation. The reversed internal AM concept \cite{VanVleck} and applications actually are communicated and applied in analysis of molecular spectra by Van Vleck in 1951 in his review article on coupling angular momenta, i.e., AM, referring to axes mounted on the molecule, adheres to opposite-sign commutator algebra.  This evolved into so-called reversed angular momentum (RAM) concepts for prediction of molecular spectra.

However, orthodox or classic QM abides by strict mathematical rules associated with the theory. Use of RAM techniques is contraindicated, especially since N\"other-type symmetry transformation \cite{ENoether1,ENoether2} sustains the standard commutator relations, viz., reversal of motion is an anti-unitary transformation, just like in the Schr\"odinger wave equation that is invariant with respect to motion-reversal or time-reversal due to anti-unitary operation, as expected. It is important to recognize that a transformation from laboratory-fixed to molecular-attached coordinates within standard QM does not condone anomalous AM operator identities \cite{Pariggerpaper}.

This review communicates proofs that the quantum-mechanic AM equations remain the same in a transition from laboratory-fixed to molecular-attached coordinates. Methods that invoke RAM for the prediction of molecular spectra are misleading. Application of standard QM establishes within the concept of line strengths \cite{Condon} consistent computation of  diatomic spectra \cite{CGPbook}; examples include hydroxyl, cyanide, and diatomic carbon spectra~\cite{CGPiramp}. First, Oscar Klein's paper \cite{OKlein} is discussed showing his original argumentation. This is followed by presenting proofs consistent with QMT opposing RAM concepts and occurrence of a minus sign in unitary and anti-unitary transformations. The ``new'' aspect of this review is the emphasis of invoking mathematics consistent with QMT. Subsequently, this review summarizes the approach for prediction of selected diatomic spectra including presentation of computed diatomic spectra of OH and C$_2$ molecules.

\section{Theory Details}
The premise of this article is Oscar Klein's work \cite{OKlein} ``Zur Frage der Quantelung des asymmetrischen Kreisels'' or ``On the question of the quantization of the asymmetric top.'' This particular work is in German without an available translation; the essential contents are in the Einleitung, viz., the introduction, and on the page following the introduction. Klein's paper reflects the initial argumentation of the RAM method, and essential aspects of this paper are discussed below,  up to Equation (\ref{RAM}).

The purpose of the 1929 work is, as O. Klein writes, to reduce quantization of the asymmetric top to simple algebra for the components of the angular momentum ``... that were developed by Dirac \cite{Dirac} and as well by Born, Heisenberg and Jordan \cite{Jordan}.'' For a solid body, the main {moments} of inertia are labeled as $A$, $B$, and $C$, the angular momenta are labeled $P$, $Q$, $R$, and one finds the CM energy of rotation, $E$,

\begin{equation}
E = \frac{1}{2}\left(\frac{P^2}{A} + \frac{Q^2}{B} + \frac{R^2}{C}\right),
\label{Kleineq1}
\end{equation}

\noindent or perhaps with convenient notation, using for operators ${\tilde J}_1 = P$, ${\tilde J}_2 = Q$, ${\tilde J}_3 = R$, where the tilde-symbol indicates that angular momenta (that would be AM operators in QM) are referred to the main axis of the ellipse of inertia (or in molecules, referred to molecular-fixed coordinates), and for {moments} of inertia $I_1 = A$, $I_2 = B$, $I_3 = C$,

\begin{equation}
E = \frac{1}{2} \sum_{k=1}^{k=3} \frac{1}{I_k} {\tilde J}_k.
\end{equation}

Subsequently, O. Klein writes that $P$, $Q$, $R$ can be understood to describe matrices satisfying QM equations of motion, with $i = \sqrt{-1}$ and using the standard $\hbar$ for Planck's constant divided by $2\pi$,

\begin{equation}
\frac{dP}{dt} = \frac{i}{\hbar}\left(EP - PE\right), \ \ \ \ \ \frac{dQ}{dt} = \frac{i}{\hbar}\left(EQ - QE\right), \ \ \ \ \ \frac{dR}{dt} = \frac{i}{\hbar}\left(ER - RE\right).
\label{Kleineq2}
\end{equation}

In terms of operators, using the Hamilton operator $\cal H$ instead of $E$ and writing the equation in the Heisenberg-picture for an abstract observable (operator), ${\cal O}$, without explicit time-dependence of the observable, i.e., $\displaystyle \frac{\partial {\cal O}}{\partial t} = 0$, and using the commutator \mbox{$\left[{\cal H}, {\cal O}\right] = {\cal H} {\cal O} - {\cal O} {\cal H}$}, gives

\begin{equation}
\frac{d{\cal O}}{dt} = \frac{i}{\hbar}\left[{\cal H}, {\cal O}\right] + \frac{\partial {\cal O}}{\partial t}.
\label{Heisenbergeq}
\end{equation}

The hypothesis of O. Klein comprises the requirement of utilizing Equation (\ref{Kleineq2}) in Equation (\ref{Kleineq1}). Consequently, O. Klein assumes commutator relations for $P$, $Q$, $R$,
\begin{equation}
i \hbar P = RQ - QR, \ \ \ \ \ i \hbar Q = PR - RP, \ \ \ \ \ i \hbar R = QP  - PQ,
\label{Kleineq3}
\end{equation}

\noindent or using abbreviated nomenclature and the Levi-Civita symbol, with $\varepsilon_{klm} = 1$ for even permutations, and $\varepsilon_{klm} = -1$ for odd ones, otherwise $\varepsilon_{klm} = 0$ for identical indices, $k, l, m = 1, 2, 3$,
\begin{equation}
\left[{\tilde J}_k, {\tilde J}_l\right] = - i \hbar \varepsilon_{klm} {\tilde J}_m.
\label{RAM}
\end{equation}

With the commutator relations in Equation (\ref{Kleineq3}), the correspondence principle leads to the equations of motion, and as O. Klein writes, ``... as we overlook occurrence of the action-quant ...,'' viz. overlook $\hbar$. Further, O. Klein remarks that Equation (\ref{Kleineq3}) differs only by the sign of $i$ from the well-known quantum-mechanical commutators for a laboratory-fixed system. In summary, O. Klein's work concludes that a minus sign is required for consistency with classical mechanics and a result of the application of the correspondence~principle.

Clearly, writing Equation (\ref{Kleineq3}) in the compact form of Equation (\ref{RAM}) highlights the minus sign that differs from the standard equations of AM operators $J_k$, $k = 1, 2, 3,$
\begin{equation}
\left[{ J}_k, { J}_l\right] = \ \ i \hbar \varepsilon_{klm} { J}_m.
\label{QMeq}
\end{equation}

The minus sign in Equation (\ref{RAM}) is labeled ``anomalous'' by some authors, e.g., J. Van Vleck~\cite{VanVleck}, but there is no justification for the anomalous minus sign to occur within QMT. Usually, one considers right-hand systems, so Equation (\ref{QMeq}) is termed as the standard quantum-mechanic AM operator identity. Sustenance of RAM concepts may appear convenient, even calling the negative sign an ``anomaly'' but without QMT support. In scientific approach and in spite of the initial success in explaining spectra within various approximations, one usually avoids starting with an ``anomaly'' and/or inaccurate presuppositions that are readily falsified \cite{Popper}. However, several textbooks and works continue support of RAM in the theory of molecular spectra \cite{refA,refB,refC,refD,refE,refF,refG,refH,refI,refJ,refK,refL,refM}, in spite of obvious falsification by QMT. This work emphasizes that there is no need to resort to RAM ``cook book'' \cite{refM} methods.

The methods in this work utilize standard QMT \cite{CohenT1,CohenT2} and standard mathematical methods \cite{Arfken}, showing that there is no sign change of the standard commutator relations when transforming from a laboratory-fixed to a molecule-attached coordinate system. Consistent application of standard AM algebra in the establishment of computed spectra yield nice agreement with laboratory experimental results \cite{CGPbook} and agreement in analysis of astrophysical C$_2$ Swan data from the white dwarf Procyon B \cite{CGPbook}, including agreement in comparisons with computed spectra that are obtained with other molecular fitting programs such as PGOPHER \cite{Western}.

Methods for measurement of optical emission signals from diatomic molecules are comprised of standard molecular spectroscopy experimental arrangements such as in laser-induced plasma or breakdown spectroscopy \cite{Kunze,Demtroder1,Demtroder2,Hertel1,Hertel2,Cremers,Miziolek,Thakur}, encountered as well in stellar plasma physics or astrophysics to name other areas of interest. Particular interests in astrophysics include ``cool'' stars, brown dwarfs, and extra-solar planets, and the associated need for accurate theoretical models for {ab initio} calculations of diatomic molecular spectra, nicely reviewed recently \cite{Tennyson}.


\section{Results}

\subsection{Angular Momentum Commutators}
The invariance of standard QMT commutator relations (see Equation (\ref{QMeq})) is communicated in this section.

\subsubsection{Invariance for Unitary Transformations}
Application of unitary transformation, viz., transforming from one coordinate system to another, leaves the AM commutator relations invariant \cite{Davydov}.
A unitary transformation operator, $U$, acting on an operator $\cal O$ $\longrightarrow$ ${\cal O}^\prime$, with $U^\dagger = U^{-1}$, is defined by
\begin{equation}
{\cal O}^\prime = U {\cal O} U^\dagger \ \ \ \ \ {\rm or} \ \ \ \ \ {\cal O} = U^\dagger {{\cal O}}^\prime U.
\label{defintion}
\end{equation}

The invariance of the AM commutators with respect to a unitary transformation, Equation (\ref{defintion}),
\begin{equation}
[J_k,J_l] = i \varepsilon_{klm} J_m \ \ \ \ \  \longrightarrow \ \ \ \ \  [J_k^\prime,J_l^\prime] = i \varepsilon_{klm} J_m^\prime,
\label{transform}
\end{equation}

\noindent can be derived by inserting $J_k = U^\dagger J_k^\prime U$ and $J_l = U^\dagger J_l^\prime U$ in Equation (\ref{transform}) to obtain the intermediate step,
\begin{equation}
U^\dagger J_k^\prime U U^\dagger J_l^\prime U - U^\dagger J_l^\prime U U^\dagger J_k^\prime U = U^\dagger J_k^\prime J_l^\prime U - U^\dagger J_l^\prime J_k^\prime U = i \varepsilon_{klm} U^\dagger J_m^\prime U.
\end{equation}

Multiplying from left with $U$ and from right with $U^{-1}$ yields the transformed identity in Equation (\ref{transform}).  In other words, a unitary transformation
preserves the quantum-mechanic AM commutators. For example, the Euler rotation matrix is easily demonstrated to be unitary \cite{CGPiramp}. In other words, there is no anomaly when going from a laboratory-fixed to a molecule-attached coordinate system.

\subsubsection{Invariance for Time Reversal or Reversal of Motion}
Time reversal or reversal of motion in QMT requires sign changes of the operators and complex conjugation, leaving the QMT commutators invariant,
\begin{equation}
[J_k,J_l] = i \varepsilon_{klm} J_m \ \ \ \ \  \longleftrightarrow \ \ \ \ \  [(-J_k),(-J_l)] = (-i) \varepsilon_{klm} (-J_m).
\end{equation}

 CM would indicate a reversal of motion when going from a laboratory-fixed to a molecular-fixed coordinate system; however, reversal of motion requires complex conjugation due to the anti-unitary requirement. In other words, the sign is preserved. QMT so-to-speak opposes the hypothesis by O. Klein.

The invariance regarding time reversal or reversal of motion of course also would apply to the abstract form of the time-dependent Schr\"odinger equation,

\begin{equation}
i \hbar \frac{\partial}{\partial t} \psi = {\cal H} \psi \ \ \ \ \ \longleftrightarrow \ \ \ \ \ (- i) \hbar \frac{\partial}{\partial (- t)} \psi = {\cal H} \psi,
\end{equation}

\noindent where $\psi$ describes an abstract vector in Hilbert space, and $\cal H$ is a Hamiltonian. Changing time $t \longrightarrow -t$ and applying conjugate complex of $i$ preserves the left-hand side of the equation. For example, for a free particle of mass $m$ and momentum $\cal P$, the Hamiltonian is ${\cal H} = {\cal P}^2 / 2m$, and the form of Schr\"odinger's equation is preserved.

Equally, the operator equation in the Heisenberg picture, see Equation (\ref{Heisenbergeq}), preserves form under time reversal or reversal of motion,
\begin{equation}
\frac{d{\cal O}}{dt} = \frac{i}{\hbar}\left[{\cal H}, {\cal O}\right] + \frac{\partial {\cal O}}{\partial t} \ \ \ \ \ \longleftrightarrow \ \ \ \ \ \frac{d(- {\cal O})}{d (- t)} = \frac{(- i)}{\hbar}\left[{\cal H}, (- {\cal O})\right] + \frac{\partial (- {\cal O})}{\partial (-t)}.
\end{equation}

 A change of sign for the operators and complex conjugation leaves the equation invariant. The mentioned symmetry can also be associated with usual N\"other symmetries~\cite{ENoether1,ENoether2}.

\subsection{Diatomic Wave Function}

For diatomic molecules, symmetry properties allow one to invoke simplifications when evaluating the laboratory wave-function in terms of rotated coordinates \cite{CGPbook}. For internuclear geometry, the spherical polar coordinates are $r$, $\phi$, and $\theta$,  and one (arbitrary) electron is described by cylindrical coordinates $\rho$, $\chi$, $\zeta$. For coordinate rotation, one uses Euler angles $\alpha$, $\beta$, $\gamma$, and without loss of generality, one can choose $\alpha = \phi$, $\beta = \theta$, $\chi = \gamma$ \cite{CGPbook}. The  result is the Wigner--Witmer eigenfunction (WWE) for diatomic molecules \cite{Wigner&Witmer1,Wigner&Witmer2},\vspace{-13pt}

\begin{equation}
\langle \rho, \zeta, \chi, \mathbf{r}_2, \dots, \mathbf{r}_N, r, \theta, \phi \, |nvJM\rangle
= \sum_{\Omega=-J}^J \langle \rho, \zeta, \mathbf{r}'_2, \dots, \mathbf{r}'_N, r \, | nv\rangle \, D_{M\Omega}^{J^{\scriptstyle*}} (\phi, \theta, \chi).
\label{WWeig}
\end{equation}

 The usual total AM quantum numbers are $J$ and $M$, and the electronic--vibrational eigenfunction is explicitly written by extracting $v$ from the collection of quantum numbers, $n$. The WWE exactly separates $\phi$, $\theta$, $\chi$. The quantum numbers $J$, $M$, $\Omega$ refer to the total AM. The sum over {$\Omega$} in Equation (\ref{WWeig}) originates from the usual abstract transformation,

\begin{equation}
|JM\rangle = \sum_{\Omega=-J}^J |J\Omega\rangle \ \langle J\Omega \, |JM\rangle,
\label{Eqabstract}
\end{equation}

\noindent where $\Omega$ is the magnetic quantum number along the rotated, or new, $z^\prime$-axis. The sum in Equation~(\ref{Eqabstract}) ensures that the quantum numbers for total AM are $J$ and $M$. In Hund's case {\it a} \cite{Bransden}, $\Omega$ describes the projection of the total AM, within L-S coupling. Hund's case {\it a} eigenfunctions form a basis; therefore, from a computational point of view, these eigenfunctions form a complete (sufficient) set. In various approximate descriptions and for specific diatomic molecules, it may be desirable to use other Hund cases.

From the rotation operator $\mathcal{R}(\alpha, \beta, \gamma)$, with the Euler angles $\alpha$, $\beta$, $\gamma$, one finds for D-matrix elements,
\begin{equation}
D_{M \Omega}^{J^{\scriptstyle*}} (\alpha, \beta, \gamma) = \langle JM | \, \mathcal{R}(\alpha, \beta, \gamma) \, | J\Omega\rangle^*.
\end{equation}

D-matrices are the usual mathematical tool for transformation from one basis to another, but the D-matrix cannot represent an eigenfunction due to presence of two magnetic quantum numbers $M$ and $\Omega$, so the sum over $\Omega$ is needed in the transformed coordinates.

Diatomic spectra composed of line positions and line strengths are based on WWE~\cite{CGPbook} instead of eigenfunctions used for the Born-Oppenheimer approximation. Extensive experimental studies confirm agreement of computed spectra with measured emission spectra from laser-induced optical plasma \cite{CGPbook}.

\subsection{Selected Diatomic Spectra}

\begin{figure}[!h]
\includegraphics[width=12.2 cm]{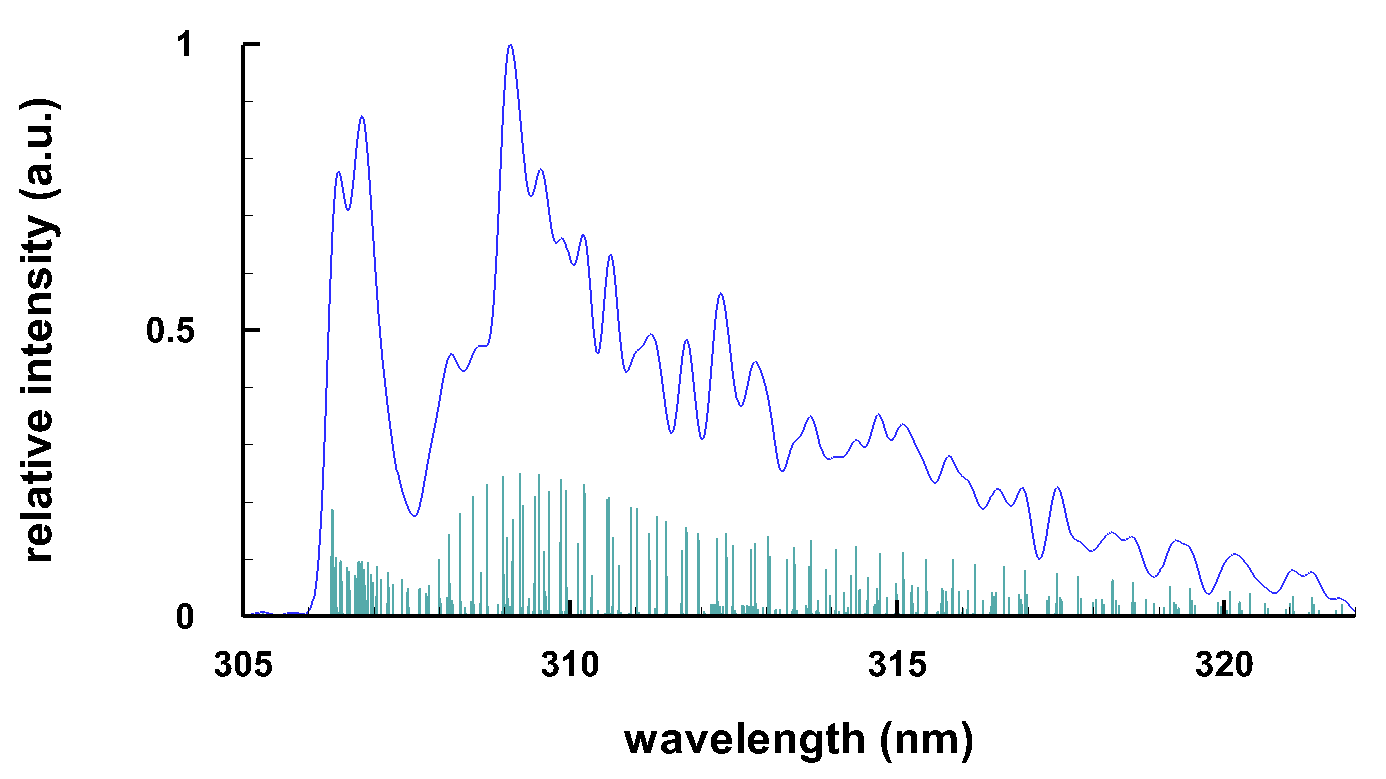}
\caption{Computed spectrum of the $A^2\Sigma \rightarrow X ^2\Pi$ uv band of OH, T = 4\,k K, (top) spectral resolutions of  $\Delta \lambda = 0.32$ nm  ($\Delta {\tilde \nu}=$ 32 cm$^{-1}$) and (bottom) idealized resolution for the stick spectrum \mbox{$\Delta \lambda = 0.002$ nm} ($\Delta {\tilde \nu}=$ 0.2 cm$^{-1}$) of the $\Delta \nu = 0$ sequence {\cite{CGPiramp}}.}
\label{figure1}
\end{figure}
\begin{figure}[!h]
\hspace{-9pt}\includegraphics[width=12.2 cm]{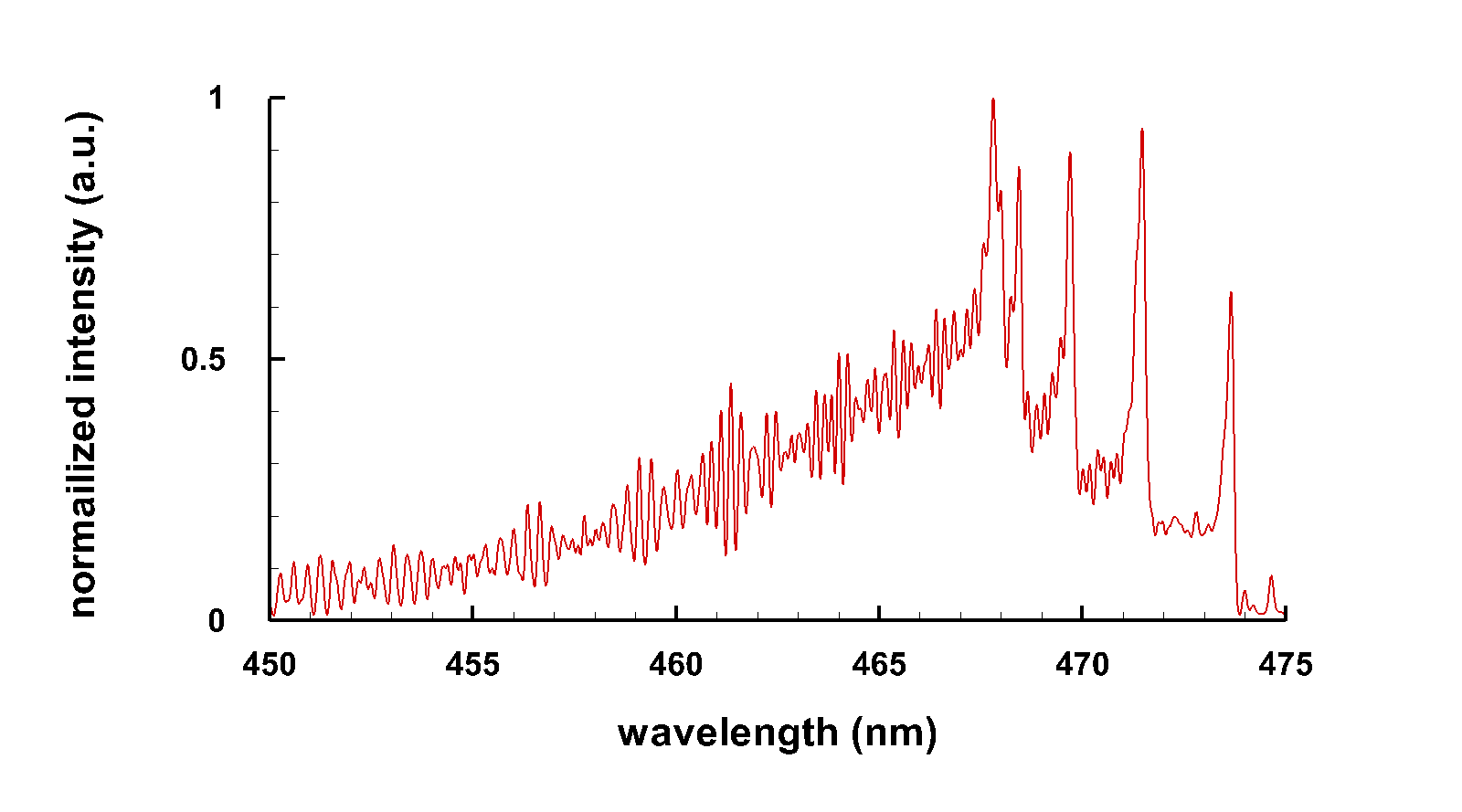}
\caption{C$_2$ Swan $d^3\Pi_g \rightarrow a^3 \Pi_u$ band $\Delta \nu = -1$ sequence, T\,=\,8\,kK, $\Delta \lambda = 0.13$ nm ($\Delta {\tilde \nu}=$ 6 cm$^{-1}$) \cite{CGPiramp}.} 
\label{figure2}
\end{figure}
\begin{figure}[!h]
\hspace{-9pt}\includegraphics[width=12.2 cm]{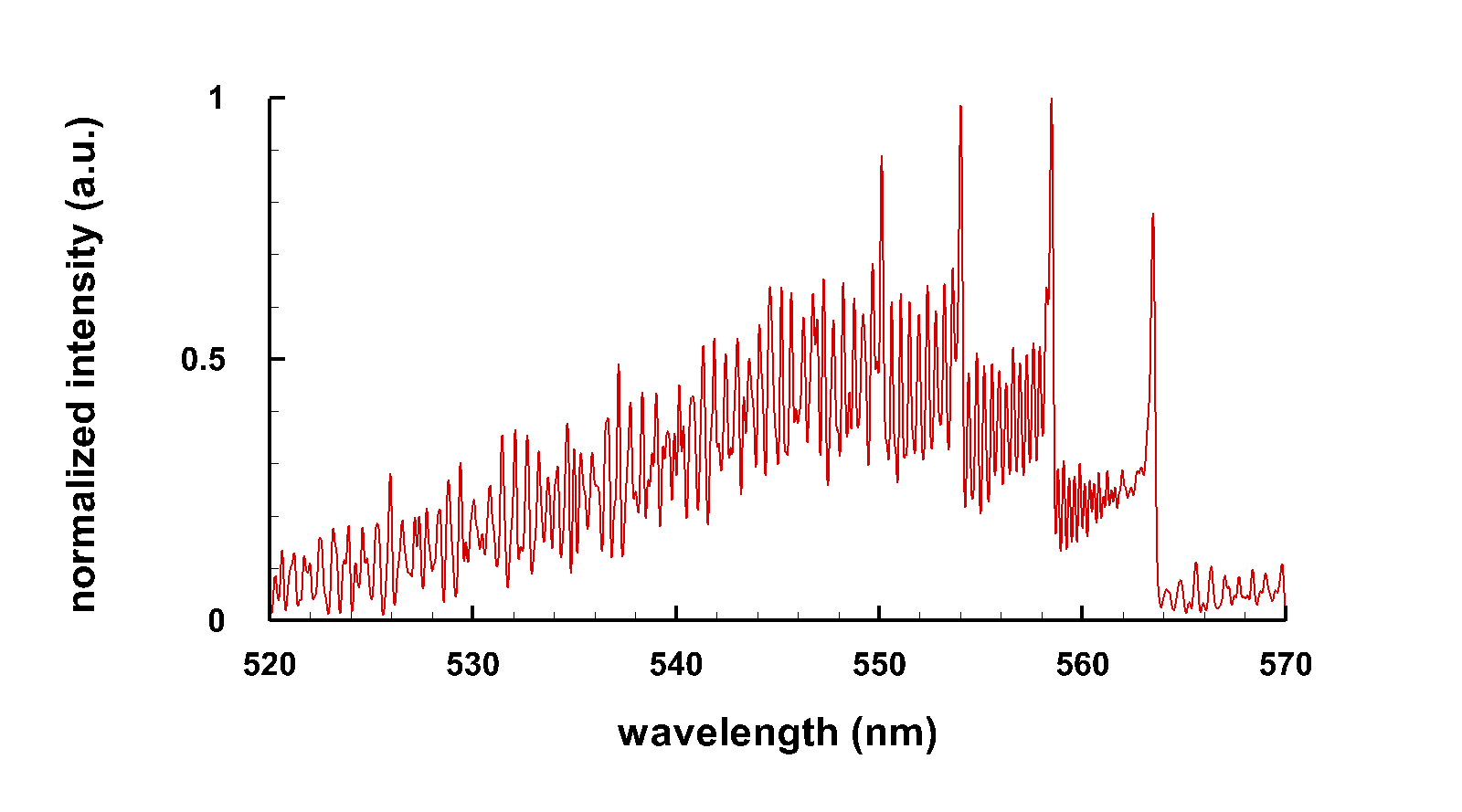}
\caption{C$_2$ Swan $d^3\Pi_g \rightarrow a^3 \Pi_u$ band $\Delta \nu = +1$ sequence, T\,=\,8\,kK, $\Delta \lambda = 0.18$ nm ($\Delta {\tilde \nu}=$ 6 cm$^{-1}$) \cite{CGPiramp}.}  
\label{figure4}
\end{figure}

Typical spectra of some diatomic molecules of general interest are presented. \mbox{Figure \ref{figure1}} illustrates OH molecular spectra for different spectral resolutions. Figures \ref{figure2} and \ref{figure4} show computed C$_2$ Swan spectra for the vibrational sequences $\Delta \nu = -1, +1$. The OH spectra, Figure \ref{figure1}, {are} a superposition of 0-0 (band head near 306 nm), 1-1 (band head near 312 nm), and 2-2 (band head near 318 nm) vibrational transition along with rotational contributions. Four C$_2$ vibrational peaks, Figures \ref{figure2} and \ref{figure4}, are clearly discernible. Rotational contributions for the selected spectral resolution, $\Delta \lambda$, appear to have beats (especially Figure \ref{figure2}) that, however, are purely coincidental.

The details for the computation, line strength data for $C_2$ Swan bands, and programs are published \cite{SABChris}. Computation of diatomic spectra utilizes high-resolution data for determination of molecular constants of selected molecular transitions from an upper to a lower energy level. Numerical solution of the Schr\"odinger equation for potentials yield r-centroids and {transition-factors} associated with vibrational transitions, viz. Franck--Condon factors. Calculated rotational factors are interpreted as selection rules because these factors are zero for forbidden transitions, viz. H\"onl--London factors. H\"onl--London factors in traditional molecular spectroscopy involve selection rules that may require use of anomalous commutators and use of two magnetic quantum numbers $M$ and $\Omega$ for a given total angular momentum $J$. Anomalous selection rules and two quantum numbers for angular momentum $J$ appear to be associated with approximations. The published line strength data \cite{SABChris,MDPIChris} are {derived consistent} with standard quantum mechanics, in other words, without anomalous commutators and without states that have two magnetic quantum numbers associated with angular momentum.

The published program package \cite{SABChris} also includes a worked high-temperature cyanide example, the Boltzmann equilibrium spectrum program (BESP) for computation of equilibrium spectra, and the Nelder--Mead temperature (NMT) routine that utilizes a non-linear fitting algorithm. The OH line strength data have been made available recently \cite{MDPIChris}.

Various reported studies of plasma spectra, including astrophysics plasma, and of molecular laser-induced breakdown spectroscopy (LIBS) \cite{SABChris,MDPIChris,PariggerLIBS,PariggerCN} illustrate nice comparisons of recorded and of computed diatomic spectra. In LIBS, plasma generated by focusing coherent radiation is analyzed primarily in visible/optical or in near-uv to near-ir regions. After initiation of optical breakdown with typically 10 nanoseconds, 100 mJ laser pulses focused in standard ambient temperature and pressure (SATP) air or in gas mixtures \cite{PariggerLIBS}, molecule formation including, for example, OH in air, C$_2$ in carbon monoxide, and CN in 1:1 molar N$_2$:CO$_2$ mixture, leads to recombination radiation that is typically measured using time-resolved optical emission laser spectroscopy. When using a metallic target, other diatomic molecules can be investigated, e.g., TiO or AlO, and molecular spectra can be computed from line strength data \cite{SABChris}.

\section{Conclusion}

Angular momentum operators are well defined in quantum mechanics theory, including the fact that there is an inherent limit in measurement of its components. Another way of formulating this could be: There are only two quantum numbers needed for description of angular momentum, usually the total angular momentum and its projection onto a quantization axis. The use of the correspondence principle to ensure compatibility with classical mechanics equations of motion brings about an {ad hoc} hypothesis of a negative sign for the commutators, as originally communicated by Oscar Klein in 1929. Subsequent application of reversed angular momentum coupling continues to find support in analytic description of molecules that also includes modeling of quantum mechanic vector-operators as vectors.

However, quantum mechanics theory already ensures how to mathematically describe angular momentum, not supporting heuristic conclusions involving reversed angular momentum concepts, nor occurrence of more than two quantum numbers for the total angular momentum of diatomic molecules. This review emphasizes that there is no mathematical justification of reversed angular momentum algebra, and it also discusses applications in diatomic molecular spectroscopy.  Consistent application of standard quantum mechanics theory is preferred, including avoidance of {a priori} use of separating electronic, vibrational, rotational wave functions. Subsequent to implementation of diatomic molecular symmetries, line strengths for selected diatomic molecules that contain effects of spin splitting and lambda-doubling as function of wavelength are in agreement with results from optical emission spectroscopy. The computed and fitted diatomic spectra nicely match within reasonable error bars, but without invoking heuristic selection rules that may be affected by initial approximations or by spurious use of reversal of angular momentum.

\section*{Acknowledgment}
{The author acknowledges support in part from the State of Tennessee funded Center for Laser Applications at the University of Tennessee Space Institute.}


\end{document}